\documentclass[authoryear,preprint,12pt]{elsarticle}
\pdfoutput=1

\usepackage{amssymb}

\newcommand{\AmS}{{\protect\the\textfont2
  A\kern-.1667em\lower.5ex\hbox{M}\kern-.125emS}}

\newcommand{\half}{\frac{1}{2}}

\newcommand{\schr}{\textrm{Schr\"odinger }}

\journal{Studies in the History and Philosophy of Science}

\begin{document}

\begin{frontmatter}



\title{Everett and the Born Rule}


\author{Alastair I. M. Rae}

\address{School of Physics and Astronomy, University
of Birmingham\\
        Birmingham B15 2TT}

\begin{abstract}
During the last ten years or so, derivations of the Born rule based
on decision theory have been proposed and developed, and it is claimed that these are valid in the context of the Everett interpretation.  This claim is critically assessed and it is shown that one of its key assumptions is a natural consequence of the principles underlying the Copenhagen interpretation, but constitutes a major additional postulate in an Everettian context.  It is further argued that the Born rule, in common with any interpretation that relates outcome likelihood to the expansion coefficients connecting the wavefunction with the eigenfunctions of the measurement operator, is incompatible with the purely unitary evolution assumed in the Everett interpretation.

\end{abstract}


\end{frontmatter}


\maketitle
\section*{Introduction}

 The conventional (``Copenhagen'') interpretation of quantum mechanics states that the result of a measurement is one (and only one) of the eigenvalues belonging to the operator representing the measurement and that, following the measurement, the wavefunction ``collapses'' to become the corresponding eigenfunction (ignoring the possibility of degeneracy).  According to the ``Born rule'', the probability of any particular outcome is proportional to the squared modulus of the scalar product of this eigenfunction with the pre-measurement wavefunction. This analysis underlies many of the predictions of quantum mechanics that have been invariably confirmed by experiment. An alternative approach to quantum measurement is the Everett interpretation (also known as the ``relative states'' or the ``many worlds'' interpretation) which was proposed by \cite{Evrt57}. The essence of this approach is that it assumes no collapse of the wavefunction associated with a measurement: instead, the time development of the state is everywhere governed by the time-dependent \schr equation. After a ``measurement-like'' event, this results in a splitting of the wavefunction into a number of branches, which are then incapable of reuniting or communicating with each other in any way. This splitting occurs even when a human observer is part of the measurement chain: the resulting branches then each contain a copy of the observer, who is completely unaware of the existence of the others.

Since its inception, the Everett interpretation has been subject to considerable criticism---e.g. \cite{Knt90}, \cite{Squr90}---which has three main strands (or branches [\emph{sic}]).  First, there is its metaphysical extravagance. The continual evolution of the universe into a ``multiverse'' containing an immense number of branches would mean that the universe we observe should be  accompanied by an immense number of parallel universes, which we do not observe and have no awareness of---surely such a postulate must be a gross breach of the principle of Occam's razor! Everett himself was aware of this criticism and, in a footnote to his original paper, he compares the conceptual difficulties of accepting his interpretation with those encountered by Copernicus when the latter proposed the (in his time revolutionary) idea that the earth moves around the sun. However, the reason that the Occam's razor argument  has not led to the universal rejection of Everett's ideas is less to do with the strength or otherwise of the Copernican analogy and more a result of the fact that the branching of the universe into the multiverse is claimed to be a direct consequence of the time-dependent \schr equation: no additional postulate, such as the collapse of the wavefunction, is required to explain the phenomenon of quantum measurement and the extravagance with universes may therefore be considered a price worth paying for the economy in postulates.

 The second strand in the criticism of Everett is known as the ``preferred basis'' problem.  This is because there is an apparent ambiguity in the way the branches are defined. Thus, if the wavefunction of a system has the form $\psi=A\psi_1+B\psi_2$, then Everett suggests that a measurement should lead to two sets of branches, one associated with each of the states represented by $\psi_1$ and $\psi_2$. However, the original state could just as well be written as $\psi=C\phi_1+D\phi_2$ where $\phi_1=2^{-1/2}(\psi_1+\psi_2)$, $\phi_2 =2^{-1/2}(\psi_1-\psi_2)$, $C=2^{-1/2}(A+B)$ and $D=2^{-1/2}(A-B)$, so why should the branches not be just as well defined by $\phi_1$ and $\phi_2$---or indeed any other orthogonal pair of linear combinations of $\psi_1$ and $\psi_2$? This problem has been largely resolved by the appreciation of the importance of the effect of the environment on a quantum system and the associated ``decoherence''---\cite{Zrk07}, \cite{Wllc02}, \cite{Wllc03a}. A quantum measurement is inevitably accompanied by complex, chaotic processes which act to pick out the particular basis defined by the eigenstates of the measurement operator. This basis is then the one ``preferred'' by the Everett interpretation and this supervenes on the \schr wavefunction. This result is now generally accepted, although \cite{Bkr06} argues that its derivation uses the Born rule so that there is a danger of circularity if it is then assumed as part of its proof.

 The third criticism leveled at Everett is the problem of probabilities.  The conventional (Copenhagen) interpretation states that, if the wavefunction before a measurement is $\psi=A\psi_1+B\psi_2$, and if $\psi_1$ and $\psi_2$ are eigenstates of the measurement operator with eigenvalues $q_1$ and $q_2$ respectively, then the outcome will be \emph{either} $q_1$ with probability $|A^2|$ \emph{or} $q_2$ with probability $|B^2|$, where these probabilities reflect the frequencies of the corresponding outcomes after a large number of similar measurements. However, according to the Everett approach there is no ``either-or'' because both outcomes are manifest, albeit in different branches. Instead of a \emph{disjunction} to which we can apply standard probability theory, we have a \emph{conjunction}, where it is hard to see how probabilities can make any sense---\cite{Squr90}, \cite{Grhm73}, \cite{Lwsd04}. There have been several attempts to resolve this conundrum and to show how probability (or something else that is in practice equivalent to it) can be used in an Everettian context. David Wallace has proposed a principle that he calls ``subjective uncertainty'' in which he claims that a rational observer should expect to emerge in one branch after a measurement, even although she is also reproduced in the other branches---\cite{Wllc03b}, \cite{Wllc07} . \cite{Grvs04} has criticized this approach and suggested an alternative in which we have to take into account the observer's ``descendants'' in all the branches, but we should ``care'' more about some than others; the extent to which we should care is quantified by a ``caring measure'' that is proportional to the corresponding Born-rule weight.  Both these approaches are designed to explain why some branches appear to be favoured over others, but both attempt to do this without altering Everett's main principle that the quantum state evolves under the influence of the time-dependent \schr  equation with nothing else added, so that the Born rule supervenes on this. An alternative approach, which I shall not discuss any further in this paper, is to maintain most of the fundamental ideas of the Everettian interpretation, but add a further layer of ``reality'' to justify the use of probabilities; an example of this can be found in \cite{Lckd89}.

 Interest in the Everett interpretation has been on the increase recently---particularly during 2007, which was the 50th anniversary of the publication of Everett's original paper, \cite{Evrt57}.  Much of the renewed interest has developed from work by \cite{Dtsh99} some eight years earlier, which was then developed by, \cite{Wllc03b},  \cite{Sndrs04} and \cite{Wllc07}.  This programme (which I refer to below by the initials DSW) aims to derive the Born rule from minimal postulates that are claimed to be consistent with the Everett interpretation, as well as with other approaches to the measurement problem.  In fact, \cite{Dtsh99} makes little reference to the Everett interpretation in his derivation of the Born rule and \cite{Sndrs04} emphasizes that he believes that his derivation is independent of any assumptions about the measurement process.  However, \cite{Wllc07} assumes the Everett interpretation and claims that his derivation shows that the Born rule is completely consistent with it. \cite{Gill05} examined Deutsch's derivation and sought to clarify the assumptions underlying it, again without referring to the Everett interpretation as such.  A similar approach, but using slightly different assumptions has been developed Zurek and is set out in a recent review paper, \cite{Zrk07}.

The present paper aims to show that some of the postulates underlying the above derivations arguments do not follow naturally from the Everett interpretation and may well not be consistent with it.

\section*{The DSW Proof of the Born Rule}

 This section sets out the DSW derivation of the Born rule by applying it to a particular example. The argument is deliberately kept as simple as possible and more general treatments can be found in the cited references.  Consider the case of a spin-half particle, initially in an eigenstate of an operator representing a component of spin in a direction in the $xz$ plane at an angle $\theta$ to the $z$ axis, passing along the $y$ axis through a Stern-Gerlach apparatus oriented to measure a spin component in the $z$ direction.

Standard quantum mechanics tells us that the initial state $\alpha_{\theta}$ can be written as a linear combination of the eigenstates of $\hat{S}_z$: $\alpha$ with eigenvalue +1 (in units of $\hbar/2$) and $\beta$ with eigenvalue -1. We have

\begin{equation} \label{eq05} \alpha_{\theta}=c\alpha+s\beta \end{equation} where $c=\cos(\theta/2)$ and $s=\sin(\theta/2)$.  Particles emerge from the two channels of the Stern-Gerlach apparatus, with the upper and lower channels indicating $S_z=+1$ and $-1$ respectively and are then detected. After they have entered the detectors, but before any collapse\footnote{At a number of points in this paper, I compare the predictions of the Everett model with those produced by the ``Copenhagen interpretation'', by which I mean a model in which the wavefunction collapses into one of the eigenstates of the measurement operator.  This is assumed to occur early enough in the process for the outcomes to be the same as would be observed if particles were emerge randomly from one or other output channel, with the relative probabilities of the two outcomes determined by the Born rule.} associated with the measurement, the total wavefunction of the system is

\begin{equation} \label{eq06} \psi=c\alpha\chi_++s\beta\chi_- \end{equation} where $\chi_+$ ($\chi_-$) is the wavefunction representing the detectors, including their environment, when a particle is detected in the positive (negative) channel. According to the Copenhagen interpretation, the corresponding probabilities for a positive or negative outcome are given by the Born rule as $c^2$ and $s^2$ respectively. From the Everettian point of view, on the other hand, there is no collapse and the system is always in a state of the form $\psi$.  However, because of the effects of the environment and decoherence, phase coherence between the two terms on the right-hand side of (\ref{eq06}) is lost, so they can never in practice interfere. The wavefunction has therefore evolved into two ``branches'' which then develop independently.

The principle of the DSW approach is to describe the process being studied as a game, or series of games, where we receive rewards, or pay penalties (i.e. receive negative rewards) depending on the outcomes. The derivation proposed by \cite{Zrk07} is quite similar to this, although it does not use game theory.

Imagine a game where the player receives a reward depending on the outcome of the experiment.  Assume that the value of $\theta$ is under our control and that, whenever the experimenter observes a particle emerging from the positive or negative channel of the Stern-Gerlach apparatus, she receives a reward equal to $x_+$ or $x_-$ respectively; these values can be chosen arbitrarily by the experimenter. In the special cases where $\theta=0$ or $\theta=\pi$, the initial spin state is an eigenstate of $\hat S_z$ with eigenvalues $+1$ and $-1$ respectively.  The particle then definitely emerges from the corresponding channel of the Stern-Gerlach apparatus and the corresponding reward is paid.

In the general case, we define the ``value''---$V(\theta)$---of the game as the minimum payment a rational player would accept not to play the game, and look for an expression for $V(\theta)$ of the form \begin{equation}\label{eq01}     V(\theta)=w_+(\theta)x_++w_-(\theta)x_- \end{equation} where the $w$s are non-negative real numbers that we call ``weights'' and which are normalized so that their total is unity. We shall find that \begin{equation}\label{eq02}     w_+(\theta)=c^2\textrm{  and  }w_-(\theta)=s^2 \end{equation} which are the probabilities predicted by the Born rule for this setup.

 First consider the effect on the wavefunction of rotating the SG magnet through $180^{\circ}$ about the $y$ axis.  It follows from the symmetry of the Stern-Gerlach apparatus that spins that were previously directed into the upper channel will now be detected in the lower channel and vice versa. Thus \begin{equation} \label{eq071} V(\theta+\pi)=w_+(\theta+\pi)x_++w_-(\theta+\pi)x_-=w_-(\theta)x_+ +w_+(\theta)x_- \end{equation} From standard quantum mechanics, the effect of this rotation on the wavefunction (\ref{eq06}) is to transform it to

\begin{equation} \label{eq07} \psi=-s\alpha\chi_++c\beta\chi_- \end{equation} We now proceed by considering a series of particular values of $\theta$.

\textbf{Case 1} The first case is where $\theta=0$ so that the initial state, $\alpha_{\theta}$, is identical with $\alpha$.  As noted above, this state is unaffected by the measurement and the particle is always detected in the positive channel.  Thus $V(0)=x_+$, $w_+(0)=1$ and $w_-(0)=0$. Similarly, $V(\pi)=x_-$, $w_+(\pi)=0$ and $w_-(\pi)=1$.

\textbf{Case 2}  In the second case, $\theta=\pi/2$ so that $\psi$ is as in (\ref{eq06}) above, but with $c=s=2^{-1/2}$. Now consider the effect of rotating the Stern Gerlach apparatus through an angle $\pi$. Using (\ref{eq071}) and (\ref{eq07}), we get the following expressions for $V$ and $\psi$ \begin{equation} \label{eq072} V(3\pi/2)=w_-(\pi/2)x_++w_+(\pi/2)x_- \end{equation} \begin{equation} \label{eq51} \psi=2^{-1/2}[-\alpha\chi_+ +\beta\chi_-] \end{equation} The only change in the wavefunction is the change of sign in the term involving $\alpha$. DSW point out that this sign, in common with any other phase factor, should not affect the value, because it can be removed by performing a unitary transformation on this part of the wavefunction only---e.g. by a rotation of the spin through $2\pi$ or by introducing an additional path length equal to half a wavelength. Moreover, \cite{Zrk07} shows that one of the effects of the interaction of the system with the environment is to remove any physical significance from these phase factors. It follows that the value should not be affected by the rotation so that $V(3\pi/2)=V(\pi/2)$, which  leads directly to

\begin{equation}\label{eq073} w_+(\pi/2)=w_-(\pi/2)=1/2 \textrm{  and  } V(\pi/2)=(x_++x_-)/2 \end{equation}

This result (which might be thought to be an inevitable consequence of symmetry) is considered by DSW to be the key point of the proof.  We should note that, although it agrees with the Born rule, it would also be consistent with any alternative weighting scheme that predicted equal weights in this symmetric situation: in particular it is consistent with a model in which the weights were assumed to be independent of $\theta$.

We now extend the result to the case where the number of output channels is $M$ instead of two and the wavefunction is the sum of $M$ terms, each of which corresponds to a different eigenstate of the measurement operator. In the case where the coefficients of this expansion are all equal, any action that has the effect of exchanging any two output channels (which are numbered 1 and 2) must leave the wavefunction unchanged apart from irrelevant changes in phase. The value is then also unchanged, but the roles of $w_1$ and $w_2$ are reversed. Hence

\begin{equation}\label{eq0731} w_1x_1+w_2x_2=w_1x_2+w_2x_1 \end{equation} where $x_i$ is the reward associated with the $i$th output channel. It follows that $w_1=w_2$; consideration of other permutations immediately extends this result to all $i$ and we have $w_i=N^{-1}$.

\textbf{Case 3} In the third case, $\theta=\pi/3$ so that $\cos(\theta/2)=\surd 3/2$ and $\sin(\theta/2)=1/2$.  We now assume that the system is modified so that, after emerging from the Stern-Gerlach magnet and before being detected, the outgoing particles interact with a separate quantum system  that can exist in one of, or a linear combination of four eigenstates $\phi_i$. Following \cite{Zrk07}, this is referred to as an ``ancilla'' from now on.  The ancilla is designed so that, if $\theta=0$ so that all spins emerge from the positive channel, the ancilla is placed in the state $3^{-1/2}\sum_{i=1,3}\phi_i$; while, if $\theta=\pi$ and all spins are negative, its state becomes $\phi_4$. From linear superposition it follows that if the original spin is in a state of the form (\ref{eq06}) with $\theta=\pi/3$, the total wavefunction of the spin plus the ancilla is

 \begin{eqnarray}\label{eq09}     \Psi=&3^{-1/2}[\phi_1+\phi_2+\phi_3]     \cos(\pi/6)\alpha +     \phi_4\sin(\pi/6)\beta\nonumber\\     =&\half[\phi_1\alpha+\phi_2\alpha     +\phi_3\alpha+\phi_4\beta] \end{eqnarray} As the coefficients of each term in the above expansion are equal, it follows from the earlier discussion of Case 2 that all four weights are equal to 0.25.  If we were to measure on the ancilla a quantity whose eigenstates were one of the functions $\phi_1$ to $\phi_4$, we should obtain a result equal to one of the corresponding eigenvalues. If the result corresponds to one of the first three eigenfunctions, we can conclude that if, instead, we had measured the spin directly, we would have got a positive result, while a result corresponding to $\phi_4$ indicates a negative spin. As this is the only such state, it follows that the weight corresponding to a negative spin is $w_-(\pi/3)=0.25$ and therefore, from normalization, that $w_+(\pi/3)=0.75$. (The last step, which follows \cite{Zrk07}, establishes these results without assuming that the weights are additive.)  The value of the game therefore equals $0.75x_++0.25x_-$.  It can also be shown quite straight forwardly---\cite{Zrk07}---that, after a number of repeats of the experiment, the predicted distribution of the results is as observed experimentally.

Following DSW, the above argument can be extended to the case of a measurement made in the absence of the ancilla if we make a further assumption, known as ``measurement neutrality''. This states that the outcome of the game is independent of the details of the measurement process---i.e. the presence or absence of the ancilla---so that $w_+(\pi/3)=0.75$ and $w_-(\pi/3)=0.25$ in either case. These quantities are identical to $\cos^2(\theta/2)$ and $\sin^2(\theta/2)$ respectively, so the derived weights are the same as those predicted by the Born rule. By choosing an appropriate ancilla, the above argument can be directly extended to examples where the ratio of the weights is any rational number, and then to the general case by assuming that the weights are continuous functions of $\theta$. Hence, the expression for the value is the same as that predicted by the Born-rule:

\begin{equation}\label{eq074}     V(\theta)=c^2x_++s^2x_- \end{equation} Further generalization to experiments with more then two possible outcomes is reasonably straightforward and does not introduce any major new principles.

\section*{Discussion}

There have been a number of criticisms of the DSW proof when applied to the Everett model in particular---e.g. \cite{Bkr06}, \cite{Bcffs00}, \cite{Lwsp05}, \cite{Lwsp07}, \cite{Hmpt07}; some of these even challenge the result (\ref{eq073}) for the symmetric case. I shall shortly develop arguments to show that, although the symmetric results appear to be consistent with the Everett model, this may not be not so in the asymmetric case.

First consider how the above translates into predictions of experimental results.  The game value is the minimum payment a rational observer would accept in order not to play the game. This means that after playing the game a number of times, a rational observer should expect to receive a set of rewards whose average is equal to the game value.  Thus, if we consider a sequence of $N$ such observations in which $n_+$ and $n_- (=N-n_+)$ particles are detected in the positive and negative channels respectively, the total reward received will be $n_+x_++n_-x_-$, and this should equal $N(w_+x_++w_-x_-)$ implying that $w_+=n_+/N$ and $w_-=n_-/N$. This, of course, is just what is observed in a typical experiment provided $N$ is large enough for statistical fluctuations to be negligible. It should be noted that frequencies are \emph{not} being used to \emph{define} probabilities, but the derived weights are used to predict the results of experimental measurement of the frequencies.

The above results are of course consistent with the standard Copenhagen interpretation, whose fundamental mantra was set out by \cite{Bohr35}: ``...there is essentially the question of an influence on the very conditions which define the possible types of predictions regarding the future behaviour of the system''.  In the present context, this means that, because an experiment designed to demonstrate interference would involve a different experimental arrangement, the experiment can be modelled as a classical stochastic system in which spins emerge from \emph{either} the positive \emph{or} the negative channel of the Stern-Gerlach apparatus. (It should be noted that this paper does not aim to justify the Copenhagen interpretation, but employs its results as a comparator with the Everettian case.)

Why should an Everettian observer have experiences such as those just described?  In the Everett interpretation, the quantum state evolves deterministically and on first sight, there would appear to be no room for uncertainty.  However, after a splitting has occurred, observers in different branches have the same memories of their state before the split, but undergo different experiences after it.  Given this, it may be meaningful for an experimenter to have an opinion about the likelihood of becoming a particular one of her successors. This introduces a form of subjective uncertainty, and \cite{Wllc07} claims that this plays a role in the Everett interpretation that is equivalent to that played by objective stochastic uncertainty in the Copenhagen case.  However, we should note that such subjective uncertainty can only come into play at the point where the experimenter becomes aware of an experimental result, in contrast to the Copenhagen model where the splitting is assumed to occur as the particles emerge from the Stern-Gerlach magnet.  I shall shortly proceed to compare and contrast the Copenhagen and Everettian interpretations of the different experiments discussed above.  To help focus the discussion, I shall initially assume that in such experiments each particular result is associated with only one branch of the final wavefunction. This assumption has been strongly criticized by DSW and others and I shall return to the question of how it affects our conclusions at a later stage.   I now analyze our earlier arguments step by step.

\textbf{Case 1: Copenhagen}  As the initial spin state is in an eigenstate of $S_z$, the result is completely determined.  The probability of the result equalling the corresponding eigenvalue is 1 and the probability of the alternative is zero.

\textbf{Case 1: Everett} There is only one branch and this contains the only copy of the observer who invariably records the appropriate eigenvalue.

There is therefore no difference between the observers' experiences in case 1 under the Copenhagen and Everettian interpretations.

\textbf{Case 2: Copenhagen}  The probabilities of positive and negative results are both 0.5.  After a large number of repeats of the experiment, the experimenter will have recorded approximately equal numbers of positive and negative results, so her average reward will be $(x_1+x_2)/2$, which is the same as the game value.

\textbf{Case 2: Everett}  The observer will split into two copies each time a spin is observed and the weights of the two branches are equal for the reasons discussed earlier. After a large number ($N$) of repeats of the experiment the vast majority of observers will have recorded close to $N/2$ positive and $N/2$ negative results and their average rewards will both equal the game value.

There is therefore no difference between the predictions of the Copenhagen and Everettian interpretations in case 2.

\textbf{Case 3: Copenhagen}  As emphasized above, this assumes that the experiment is a stochastic process in which a particle emerges from either the positive and or the negative channel and the relative probabilities of the outcomes are equal to the Born weights. In the presence of the ancilla, a particle is detected in one (and only one) of the equally-weighted states $\phi_1$ to $\phi_4$, and all four outcomes have equal probability. To have been observed in any of the first three states, the spin must have emerged from the Stern-Gerlach experiment through the positive channel, while if the final result corresponded to $\phi_4$, it must have come through the negative channel. It follows directly that if the ancilla were absent, three times as many spins would be detected as positive than as negative.  Thus, the principle of measurement neutrality, assumed in stage 3 of the earlier derivation, follows naturally from the assumptions underlying the Copenhagen interpretation.

 \textbf{Case 3: Everett}   We first consider the situation where an ancilla is present so that the state is described by (\ref{eq09}); there are therefore four equally-weighted branches, one corresponding to each of the $\phi_i$.  The observer splits into four equally-weighted copies and should expect her descendants to record an equal number of each of the four possible results and therefore conclude that there are three times as many positive as negative spins. However, in the absence of an ancilla, there are only two branches and the observer is split into two copies each time a result is obtained. To show that a typical Everettian observer should record results that are consistent with the Born weights, we again have to apply the principle of measurement neutrality. We saw above that this is a natural, if not inevitable, consequence of the Copenhagen interpretation, but we shall now demonstrate that this is not the case in an Everettian context.

Under the Copenhagen interpretation, particles are assumed to emerge from \emph{either} the positive \emph{or} the negative channel and then into one, \emph{and only one}, of the states $\phi_i$. This is not true in the case of the Everett interpretation, where the system evolves deterministically and the state is described by a linear combination of the wavefunctions associated with a particle being present in each channel. Apparent stochasticity, or subjective uncertainty, only enters the situation at the point where the experimenter observes the result and splits into a number of descendants---two in the absence of the ancilla and four if it is present.  There is no requirement for the frequencies to be the same in both cases---i.e. no \emph{a priori} reason to apply the principle of measurement neutrality.  In the language of decision theory, the values of the two games are not necessarily the same, so a decision on whether or not to accept a payoff may depend on whether the game is being played with or without an ancilla.  Indeed, as in the absence of an ancilla there are only two branches, we might expect each observer's experience to be the same as in case 2, with equal numbers of positive and negative results and an equal reward for each outcome---i.e. the statistical outcomes would be independent of the weights. I shall argue later that this is a natural consequence of the Everettian interpretation, but at present simply emphasize that the principle of measurement neutrality is a self-evident consequence of the assumptions underlying the Copenhagen interpretation, but constitutes a major additional postulate in the context of Everett.

\begin{figure} \begin{center}     \includegraphics[width=10cm]{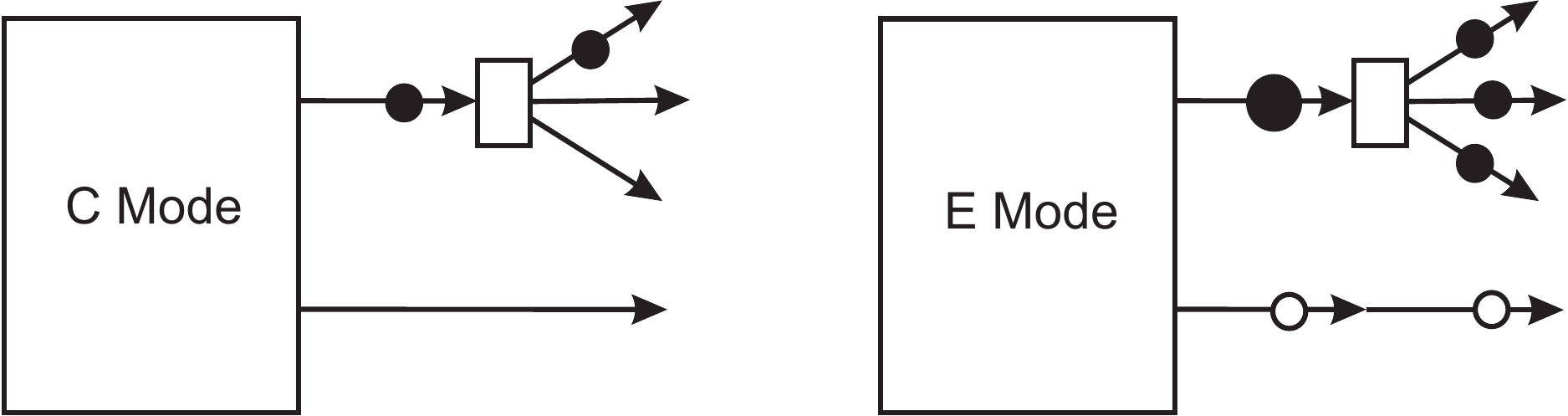} \caption{\small In the cop mode a ball is emitted from the first box through either the upper or the lower port and detected either before or after entering the second box; the figure shows one possible outcome.  In the eve mode, balls emerge from both ports and one of them is detected either before or after the second box, which releases three balls every time one enters.}   \label{fig1} \end{center} \end{figure}

I further illustrate this last point by considering a simple classical example that consists of a box with two exit ports from each of which a series of balls emerges as in figure \ref{fig1}. The apparatus can be operated in one of two modes that we denote as ``C'' and ``E''. In the C mode, balls emerge one at a time from one of two output ports and, on average, three times as many come out of the upper port as from the lower. An experimenter observes the balls as they emerge and confirms this relative likelihood. Still in the C mode, the experiment is modified so that when a ball emerges from the upper port, it passes into a second, ``ancillary'' box and then emerges at random through one of three output channels before being detected. The experimenter now detects a ball either in one of these three channels or emerging from the lower port. Clearly the first of these results is three times as likely as the second, so the observed frequencies are independent of the presence or absence of the second box.  Thus the equivalent of measurement neutrality holds in this case.

Now consider the game in the E mode, which is also illustrated in figure \ref{fig1}. In this case two balls emerge from the box simultaneously: a black ball, from the upper port and a white ball from the lower. The two balls fall into a receptacle (not shown in the figure) and an experimenter draws one at random; after repeating the experiment a number of times she sees equal numbers of black and white balls. The experiment is now modified so that the black balls are directed into an ancillary box which now contains a device that releases three identical black balls, one through each of the three output ports, whenever one enters. These three balls along with the white one now fall into the receptacle and the observer again draws one at random; she now sees a black ball three times as often as a white ball. Thus, the relative likelihood of a black or white ball depends on the presence or absence of the second box, and we can conclude that measurement neutrality is not necessarily preserved when the game is played in the E mode.

 A more whimsical analogy follows the precedent set by Schr\"odinger's cat by using animals to illustrate our point. First consider Copenhagen rabbits. These come in two colours---black and white; they  are all female and capable of giving birth to one (and only one) baby rabbit which is always of the same colour as its mother. Let us suppose we have four Copenhagen rabbits, three black and one white in a hat and suppose that one, of them, chosen at random, is pregnant.  We first play the game of ``pick out the pregnant rabbit'' by putting our hand in the hat, identifying and then pulling out the pregnant rabbit.  We are paid different rewards ($x_b$ and $x_w)$ depending on whether the extracted rabbit is black or white.  After playing the game a number of times, we find that we have pulled out three times as many black rabbits as white, so that the game value is $(3x_b+x_w)/4$. The second game is one where we wait until the pregnant rabbit has given birth and then pull out and identify the colour of the baby.  Clearly the results and the value are the same as in the first game.

Now consider Everettian rabbits, which are also either black or white. In contrast to the Copenhagen rabbits, they are capable of carrying and giving birth to more than one offspring. Suppose we have two pregnant Everettian rabbits:  a white rabbit that is pregnant with a single offspring, and a black rabbit that is expecting triplets. If we draw one of the two pregnant rabbits from the hat at random, the game value will be $(x_b+x_w)/2$. However, if, instead, we wait until after the rabbits have given birth and then draw out one of the offspring at random, the game value will now be $(3x_b+x_w)/4$. Thus Copenhagen rabbits preserve measurement neutrality, but Everettian rabbits do not.

Given the assumptions underlying the Copenhagen interpretation, the first game in the C mode and the game with the Copenhagen rabbits form close parallels with the quantum example discussed earlier. Similarly, the first game in the E mode and the game with Everettian rabbits are closely parallel to the quantum case, provided we accept that random selection at the point where the observer becomes aware of the result is equivalent to subjective uncertainty in the quantum case.

In both these examples as well as in the quantum case, I have shown that measurement neutrality is not a necessary consequence of the principles underlying the Everett interpretation. However, in all the cases where it need not apply, the symmetry is broken in the sense that the weights associated with the different outcomes are not equal.  It follows that measurement neutrality (or, indeed, some other quite different principle) could be restored in the classical examples by making additional assumptions: for example, it could be arranged that the ball emerging from the upper channel in the E game is heavier than that coming out of the lower, and that it is three times easier to find and extract a more massive ball when making the selection; similarly, it might be three times easier to catch a rabbit carrying triplets than one pregnant with a single offspring. However, such \emph{ad hoc} rules would have to be built into the physics of the set up when it was designed and constructed. In the quantum case under the Everett interpretation, measurement neutrality therefore has to be an additional assumption, rather than following directly from the structure of the theory as in the Copenhagen case.  \cite{Gill05} shows that measurement neutrality is equivalent to assuming that measures of probability are invariant under functional transformations ---i.e. the probability of obtaining a particular result when measuring a variable is the same as that pertaining when a function of the variable is measured.  He considers that functional invariance in the case of one-to-one transformations is ``more or less definitional'', but is much less obvious in the many-to-one case, which is required for situations such as case 3 above.  Gill's discussion relates to probabilities as conventionally defined and his paper makes no reference to the Everett interpretation.  I believe that the above argument shows that many-to-one transformations are also ``more or less definitional'' under the Copenhagen interpretation, but not in the Everettian context.

 Measurement neutrality and an associated principle that he calls ``equivalence'' have been argued for by David Wallace in a number of papers---\cite{Wllc02}, \cite{Wllc03a},\cite{Wllc03b}, \cite{Wllc07}. He considers games in which the measurement result is erased after it triggers an associated reward and before the experimenter has recorded the outcome.  In the symmetric ($\theta=\pi/2$) case, the final states are independent of the pattern of rewards, which reinforces the arguments leading to (\ref{eq073}). However, this is not an issue in the present discussion, which challenges the assumption of measurement neutrality only in  the asymmetric case.   Another point emphasized by \cite{Wllc07} is that the boundary between what is usually taken as preparation and what is part of the ``actual'' measurement is essentially arbitrary, particularly in the context of the Everett interpretation.  However, the observation and recording of the result by a conscious observer is part of the measurement proper, and it is only at this point that subjective uncertainty or the relevance of a caring measure is introduced into the Everettian treatment of the Born rule.

Up to this point I have argued that the assumptions underlying the derivation of the Born rule, in particular measurement neutrality, are not necessary in an Everettian context, though they may be treated as added postulates.  I now intend to go further and argue that there is an inconsistency between the assumptions underlying the Everett interpretation and the Born rule---or, indeed any rule that relates the likelihood of a measurement outcome to the amplitudes ($c$ and $s$ in the above example) associated with the branching of the wavefunction in a non-trivial way.  I shall continue to use the example of the measurement of the spin component of a spin-half particle as a focus of the discussion.

The scenario I now discuss is one where an observer (``Bob'') records the number of positive spins ($M$) in a set of measurements of the state of $N$ identically prepared spins that have passed through a Stern-Gerlach apparatus.  We consider the particular case where Bob \emph{does not know the value of $\theta$ before he makes any measurements}; that is, he has not seen the apparatus or been told how the magnet is oriented, which means that his initial state is represented by a wavefunction which is independent of $\theta$.   However, if Bob knows the Born rule, he can estimate the value of $\theta$ as $2\cos^{-1}(M/N)^{1/2}$ and his confidence in this value will be the greater, the larger are $M$ and $N$.  As a result of this experience, Bob's state has been changed from one of ignorance to one where he has some knowledge of $\theta$. This change must therefore have been reflected in Bob's quantum state, causing a modification to his wavefunction, which now depends on $\theta$. To further emphasize this point, suppose that the value of $\theta$ can be changed without Bob's direct knowledge by another experimenter (``Alice'') who has control of the Stern-Gerlach apparatus. If she does this and the experiment is repeated a number of times at the new setting, Bob will find that his expectations have been consistently wrong. He may initially attribute this to statistical fluctuation, but eventually he will amend his state of expectation to bring it into line with his experience.  Indeed, Bob may know that Alice is able to do this, in which case he will be more likely to amend his state of expectation at an earlier stage. Alice could then send signals to Bob by transmitting sets of $N$ particles using the same value of $\theta$ for each set, but changing it between sets. If the Born rule applies, Bob can deduce the values of $\theta$ that Alice has used from the relative numbers of positive and negative results, so Alice has again caused changes in Bob's state of expectation and therefore of his wavefunction.

It is one of the principles of the Everett interpretation that, once branching has occurred and the possibility of interference between branches has been eliminated, the wavefunction associated with a branch describes the ``relative state'' of the system contained in that branch, which cannot be influenced by the state of any other branch.  Moreover, the form of the relative state functions, which represent the whole branch including the version of Bob associated with it, are the same whatever are the values of the expansion coefficients $c$ and $s$.   This implies that the properties of a system represented by such a relative state are not affected by the measuring process. Thus, although these constants enter the expressions, they do so only as expansion coefficients, which have no effect on the wavefunctions of the relative states associated with the component branches.   In particular, the observer's state of knowledge of the value of $\theta$ cannot be altered as a result of this process.  This is in direct contradiction to the conclusion reached above, assuming that the Born rule holds.  There is therefore an inconsistency between the principles underlying the Everett interpretation and the appearance of a correlation between the apparatus setting and the relative frequencies of the possible outcomes, such as is implied by the Born rule.

To develop this point further, consider the state of the whole system after $N$ particles have passed through the apparatus, so that, according to the Everett interpretation, the wavefunction contains $2^N$ branches that correspond to all possible sequences of the results of the measurements performed so far.  That is, using (\ref{eq06}),

\begin{equation} \label{eq061}
\prod_{i=1,N}\alpha_{\theta}(i)\chi_0\longrightarrow \sum_{P_{s_i}}c^ms^{N-m}\Psi(s_1,s_2,...s_N)
\end{equation}
where $\alpha_{\theta}(i)$ is the initial state of spin $i$ and $\chi_0$ refers to the initial state of the detecting apparatus, including the observer Bob, which is independent of $\theta$, given the assumptions set out earlier.  Each parameter $s_i$ has two values, $+$ and $-$; $\Psi(s_1,s_2,...s_N)$ represents the state of the whole system (i.e. spins, measuring apparatus and Bob) after the results $s_i$ have been recorded in a measurements on spin $i$ for all $i$ from $1$ to $N$; $m$ equals the number of positive spins in this set;  $\sum_{P_{s_i}}$ implies a summation over all $2^N$ permutations of $s_i$.  Each term in the summation refers to a separate branch in the Everett interpretation.

It follows from (\ref{eq061}) that the number of branches in which $m$ positive results have been recorded is $N!/m!(N-m)!$ and the Born weight associated with this whole subset equals $c^{2m}s^{2(N-m)}$.  Under the Copenhagen interpretation, the probability of observing $m$ positive results is the product of these two quantities: this has a maximum value when $m=M=Nc^2$ ($=3N/4$, if $\theta=\pi/3$ as in case 3 above) and a standard deviation of $|cs|N^{1/2}$ ($=\surd 3/4$). Suppose now that the Everett assumptions hold so that there has been no collapse.  After the measurement, the wavefunction (\ref{eq061}) will consist of a linear combination of branches, each of which contains a version of Bob who has recorded a value for $m$.  If $N$ is large, the vast majority of observers will observe approximately equal numbers of positive and negative results and a small minority will observe results in the vicinity of the ratio predicted by the Born rule.  Repeating the experiment with a different value of $\theta$ does not change the number of observers recording any particular result, so, if this were all there were to it, Bob's experience would not correlate with the apparatus setting and he would be unable to  deduce a reliable value of $\theta$ from his observations.  However, the Everett interpretation only works if this is not all there is to it.  Because of subjective uncertainty, an observer's successors in branches that have a high Born-rule weight are somehow favoured over the others.  How this can work is at the heart of the difficulties many critics have with the Everett interpretation, but let us leave this on one side.  The fact that these successors are so preferred, means that they can with confidence deduce the value of $\theta$ from their observations of $M$ and $N$.  Acquiring this information must therefore have altered their reduced state, in contradiction to the Everettian assumptions set out above.

Several points should be noted about the above. First, the contradiction does not arise in the Copenhagen interpretation because, as noted earlier, this assumes that stochasticity arises at the point where the spin emerges from the Stern-Gerlach magnet.  The information as to which branch is occupied by the spin is additional to that contained in the wavefunction and is obtained by Bob through the collapse process.  Hence, no contradiction arises when this is used by the experimenter to guide his expectations about subsequent measurements.

Second, it should be emphasized that the argument applies only to information about the apparatus setting that is obtained by Bob \emph{as a result of the measurement process}.   He could of course have been told in advance how the apparatus was set up so, in this case, $\chi_0$ would already be a function of $\theta$.  The later argument could probably be extended to show that he should not be able to obtain further information about $\theta$ by the measurement process, but I believe it clarifies the discussion if we focus on the case where Bob has no prior knowledge of $\theta$: to demonstrate inconsistency, it is only necessary to establish a contradiction in at least one particular case.

Third, although I have focussed on the Born rule, the above arguments would apply equally well to any model in which the outcome frequencies were assumed to depend systematically on the expansion coefficients.  This is of rather marginal interest given that the Born rule is the one that is established by experiment.

If we accept the above, it follows that the only way probability should be able to enter the Everett interpretation is if all branches are assigned equal weight. Might it nevertheless be possible to reconcile this conclusion with experiment?  Up to now, we have assumed one branch per outcome, without attempting to justify this.  We now turn to the question of ``branch counting'', which means considering the number of branches associated with any given measurement outcome. If we accept the argument that the expansion coefficients play no role in determining the outcome likelihood in an Everettian context, then an experimenter's expectation of a particular outcome should be proportional to the number of branches associated with it. Such an assumption is similar to that made in statistical thermodynamics, where the ergodic hypothesis states that the result of averaging over an ensemble of systems is the same as the time average for a single system.  When applied to the symmetric case, this is an essential part of the arguments leading to (\ref{eq073}) and (\ref{eq0731}).  However, branch counting has been strongly criticized by DSW on a number of grounds. \cite{Wllc07} considers a scenario in which extra branching is introduced into one (say the plus) channel by associating with it a device that displays one of, say, a million random numbers. He argues that this must be irrelevant to an experimenter who sees only the measurement result and is indifferent to the outcome of the randomizing apparatus. This is because ``if we divide one outcome into equally-valued suboutcomes, that division is not decision-theoretically relevant''. However, this argument does not fully take into account the Everettian context. Referring again to the classical game discussed earlier and illustrated in figure \ref{fig1}, we can consider the additional branching on the right of both setups as due to the presence of a randomizer with three possible outputs.  In the case of the C game, these are indeed irrelevant to the expectation of the player, because a ball emerges from only one of the three channels and must therefore have passed through the upper channel at the previous stage.  However, in the case of the E game, the chances of observing a black ball are enhanced (tripled) by the splitting and this would have to be taken into account by any rational player, even if the only result she sees is the colour of the ball. Similarly, if we introduce a random number machine as Wallace suggests, then its state will be a linear combination of its million possible outcomes and all these will be associated with a positive value of spin. Given that there is only one branch associated with the alternative outcome, we could well expect the subjective likelihood of a positive result to be one million times greater than that for a negative outcome.

A second argument deployed to criticize branch counting is based on the fact that the interaction of a quantum system with its environment leads to an immensely complex branching structure. Indeed it is claimed by DSW that the number of branches is not only very large (possibly infinite), but is also subject to very large and rapid fluctuations before, during and after the observation of a result; which may mean that it is not meaningful to talk about even the approximate number of branches that exist at any time. This is adduced as a reason why a rational player should ignore the complexity of the branching structure and instead expect to observe results consistent with the Born rule. However, if the likelihood of observing a particular result is proportional to the number of associated branches, the complexity introduced by decoherence should actually result in the outcome of a measurement being completely unpredictable. The situation is similar to chaos in classical mechanics or to turbulence in hydrodynamics, whose onset certainly does not lead to increased predictability.  In the arguments above, we assumed that each outcome was associated with a single branch, so what would be the likely consequences of a complex branch structure in an Everettian context? First, there may well be situations in which we could expect the number of branches associated with different outcomes to be equal, at least when averaged over a number of measurements, and in this case our earlier discussion would not be affected. However, we might be able to devise a situation (e.g. one in which a detector was placed in the positive output channel only) where the numbers of branches in the two channels would be expected to differ greatly. We should then expect to detect a larger number of (say) positive than negative results This would be true even if the Stern-Gerlach apparatus were oriented symmetrically---i.e. with $\theta=\pi/2$, so the symmetry on which we based some of our earlier arguments would not hold.  The complexity and fluctuations of the branch structure in the Everett case would render even the statistical results of a quantum measurement unpredictable.  Such a situation is sometimes described as being ``incoherent'' and it has been argued that this would mean that the universe would be nothing like the one we experience. However, the obvious conclusion to draw from this is that the Everett assumptions are falsified, rather than that the Everett model is correct and the arguments based on it that lead to this incoherence must be wrong.

It might be thought that branch counting could restore the Born rule if the number of branches associated with a particular outcome were proportional to the Born weight.  However, not only is there no obvious mechanism to achieve this, it is also inconsistent with the Everett model for the same reasons as were set out earlier.  The quantum description of the branch structure is contained within  $\Psi$ in (\ref{eq061}) and therefore cannot depend on the expansion coefficients for the reasons argued above.

\section*{Conclusions}

I have argued that attempts to prove the Born rule make assumptions that are essentially self-evident in the context of the Copenhagen interpretation, but not with the Everett model of measurement.   I have further argued that probabilities which are functions of the expansion coefficients are not consistent with the Everett interpretation, because these quantities are not then accessible to an observer in the reduced state associated with a branch.  An alternative scheme that could be consistent with Everett is one where each branch has the same probability and the probability of a given outcome depends on the number of branches associated with it.   However, this also cannot be made consistent with the Born rule and it leads to predictions of chaotic, unpredictable behaviour, in contrast to the relatively well-ordered behaviour, invariably demonstrated in experiments.  I conclude that the Born rule is a vitally important principle in determining quantum behaviour, but that it depends on wavefunction collapse, or something very like it, that does not supervene upon the time-dependent \schr  equation.    It would be possible to retain the many-worlds ontology of the Everett model while allowing information to be transferred through the measurement, but the state evolution would no longer be governed by the \schr equation alone and the economy of postulates would no longer obviously outweigh the metaphysical extravagance associated with the Everett picture.

The debate between the different interpretations of quantum mechanics has often been metaphysical in the sense that they often make the same predictions and cannot therefore be be distinguished experimentally. The present paper has argued that this is not so in the case of the Everett interpretation, which predicts results different from those that follow from the Copenhagen interpretation, which in turn are supported by experiment.  If this is accepted, the Everett model will have been falsified and the search for a consensual resolution of the quantum measurement problem will have to be focussed elsewhere.

I am grateful to Peter Lewis for comments on an earlier draft of this paper and to Simon Saunders and David Wallace for useful discussions.  I should also like to thank the referees for comments that have helped me clarify some of the points made in the discussion.







\end{document}